\documentclass[reprint,amsmath,amssymb,aps,prl]{revtex4-1}

\usepackage{graphicx}
\usepackage{color}
\usepackage{multirow}
\usepackage{bm}
\usepackage[breaklinks=true,colorlinks,citecolor=blue,linkcolor=blue,urlcolor=blue]{hyperref}

\def\be {\begin{equation}}
\def\ee {\end{equation}}
\def\bea{\begin{eqnarray}}
\def\eea{\end{eqnarray}}

\begin{document}

\title{Linear Magnetoresistance as a Probe of the N\'eel Vector in Altermagnets with Vanishing Anomalous Hall Effect}
\author{Kamal Das}
\author{Binghai Yan}
\affiliation{Department of Condensed Matter Physics, Weizmann Institute of Science, Rehovot 7610001, Israel}
\affiliation{Department of Physics, The Pennsylvania State University, University Park, Pennsylvania 16802, United States}

\begin{abstract} 
Despite time-reversal breaking in momentum space, several altermagnets remain electrically silent to the primary characterization tool anomalous Hall effect, due to crystalline symmetries, jeopardizing their experimental identification. 
Here, we show that time-reversal odd magnetoresistance exhibiting butterfly-like hysteresis with linear magnetic field dependence near the zero field provides a robust transport signature of altermagnetism even when the anomalous Hall effect vanishes. Using semiclassical theory and symmetry analysis, we demonstrate that this effect is generic across altermagnets and validate it through first-principles calculations in CrSb. Our results establish linear magnetoresistance as an alternative detection of the Berry curvature and Néel order in unconventional antiferromagnets. 
\end{abstract}

\maketitle

{\it Introduction---} Altermagnets (AMs) are a newly identified class of compensated collinear magnets that host momentum-dependent spin splitting despite vanishing net magnetization~\cite{libor_SA2020_crystal,Yuan2020,ma2021multifunctional,hayami_JPSJ2019_momentum,libor_PRX2022_beyond,libor_PRX2022_emerging,Mazin2022,bai_AFM2024_alter, song_NRP2025_alter}. This unusual combination of antiferromagnetic order and spin-split electronic bands makes them promising for future spintronics applications~\cite{naka_NC2019_spin,shao_NC2021_spin,libor_PRX2022_giant}.

The anomalous Hall effect (AHE) is widely regarded as a signature of time reversal $({\mathcal T})$-symmetry breaking in AMs~\cite{libor_SA2020_crystal,gonzalelz_PRL2023_spon,reichlova_NC2024_obser,zhou_N2025_mani,zhou_arxiv2026_surfaces,zhou_N2025_mani}  along with the spin split revealed by angle-resolved photoemission spectroscopy (ARPES)~\cite{song_NRP2025_alter}. However, AHE vanishes in many AM materials, including RuO$_2$~\cite{berlijn_PRL2017_itinerant,zhu_PRL2019_anomalous,hiraishi_PRL2024_nonmag,keler_npjs2024_absence}, CrSb~\cite{reimers_NC2024_direct,yang_NC2025_three,ding_PRL2024_large,zeng_AS2024_observa},  KV$_2$Se$_2$O~\cite{jiang_NP2025_a} and Ca$_3$Ru$_2$O$_7$~\cite{mali_NC2026_probing,zhao_PRB2025_nonlinear}, because of related crystal symmetries. As a result, these systems are electrically ``silent" to this primary characterization tool, obscuring the detection of their underlying AM order.

\begin{figure*}[t!]
\centering
\includegraphics[width =\linewidth]{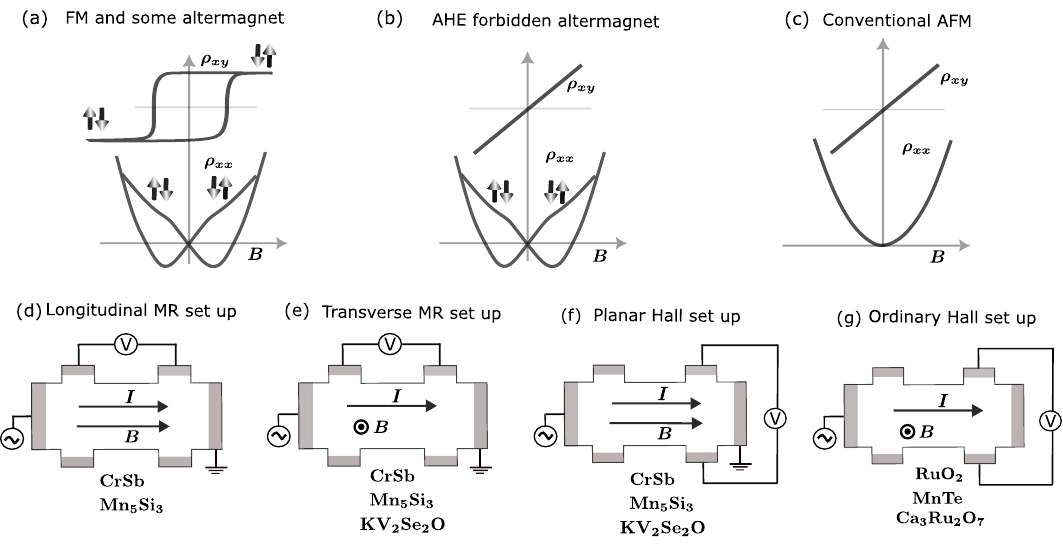}
\caption{(a) Typical anomalous Hall effect (AHE) and linear MR in butterfly-like hysteresis loop observed in ferromagnet and in some altermagnets (AMs). (b) When crystal symmetry forbids the AHE, the linear MR can still appear in the hysteresis loop, reflecting the N\'eel order orientation. (c) In the conventional AFMs with ${\mathcal P}{\mathcal T}$ and ${t} {\mathcal T}$ symmetry, the linear response measurement shows no hysteresis and hence can not detect N\'eel order. (d)-(g) The different linear MR measurements set up for the various AM candidates. (d) The longitudinal MR set up when all the fields are in the same direction. (e) The transverse MR set up when the magnetic field is perpendicular to the current. (f) The planar Hall set up, where the magnetic field is confined in the 2D plane of measurement. (g) The ordinary Hall set up where all the fields are perpendicular. The corresponding AM materials are indicated below each configuration. For Mn$_5$Si$_3$ and MnTe, spins in the hexagonal plane and along $2 \bar 1 \bar 1 0$ axes are considered, respectively.}
\label{fig_1}
\end{figure*}

In this work, we propose to use a linear magnetoresistance (MR) to detect the AM order even when AHE vanishes. In known ferromagnetic materials, the AHE comes together with a linear MR characterized by the butterfly-like hysteresis~\cite{west_JAP1961_magneto,shtrikman_SSC1965_remarks,xiao_PRB2020_linear,wang_NC2020_anti,sunko_PRB2025_linear,vorobev_PRB2024_d, farajollahpour_npjQM2025_berry}, as illustrated by Fig.~\ref{fig_1}(a). Thus, this linear MR is commonly presumed to be a relevant feature of AHE. Here, we point out that the linear MR generally exist for AMs, distinguishing them from conventional antiferromagnets (AFMs) which have either parity-time reversal (${\mathcal P} {\mathcal T}$) or translation-time reversal (${t} {\mathcal T}$) symmetry, regardless whether AHE appears or not. Because it is odd to time-reversal, this linear MR changes sign upon reversing the N\'eel vector of AM.
Through symmetry analysis and first-principles calculations, we demonstrate the linear MR in recent experimentally realized AMs, which exhibit both longitudinal (e.g., $R_{xx}$) and transverse (e.g., $R_{xy}$) components. For example, we predict a linear MR of $\sim 0.1 \%$ in CrSb at $B=3$ T. Our results provide an alternative way to identify the altermangetism in AHE-absent materials.

{\it Symmetry and phenomenology of ${\mathcal T}$-odd linear MR---} In ordinary materials, the Onsager's relation requires $R_{xx}({\bm B})=R_{xx} (-{\bm B})$ and therefore the leading term of MR is $B^2$. 
In magnetic materials, however, the Onsager's relation has the modified form 
$ R_{ij}({\bm B},{\bm M})= R_{ji} (-{\bm B},-{\bm M})$
in the linear-response regime, in which we denote the ferromagnetic or antiferromagnetic order parameter by ${\bm M}$. 
In this case, the $B$-odd terms are allowed in MR. In the perturbative regime with a small ${\bm B}$ field, when ${\bm M}$ is independent of ${\bm B}$, 
the resistance has a linear-$B$ correction, $ R_{ij}({\bm B},{\bm M})= R_{ij}^0({\bm M}) + R_{ij}^l ({\bm M}) B_l + \cdots $ and satisfies~\cite{shtrikman_SSC1965_remarks,xiao_PRB2020_linear}
\be
R_{ij}^{l}({\bm M}) = - R_{ji}^{l} (-{\bm M})~.
\ee
This sign change under the magnetic order reversal manifests in the butterfly-like loop of resistance, as shown in Fig.~\ref{fig_1}(a).

The sign reversal of linear MR is reminiscent of the AHE, but with an important distinction. While AHE is described by a second-rank tensor ($R_{ij}$), the linear MR arises from a third-rank tensor ($R_{ij}^l({\bm M})$) that couples the electric field and the axial magnetic field. As a result, this effect is less symmetry-constrained and can exist even when AHE vanishes~\cite{xiao_PRB2020_linear,sunko_PRB2025_linear} as highlighted in Fig.~\ref{fig_1}(b). For example, in mirror-symmetric AMs such as RuO$_2$, Ca$_3$Ru$_2$O$_7$, or MnTe (spins $\parallel 2\bar 1\bar 10$)~\cite{amin2024nanoscale,robler_arxiv2025_low}, mirrors suppress AHE, yet linear MR remains allowed. Likewise, in thin-film Mn$_5$Si$_3$~\cite{reichlova_NC2024_obser,badura_NC2025_obser}, ${\mathcal C}_{2z}{\mathcal T}$ symmetry forbids AHE for in-plane spin orientations, a linear MR is still permitted. Other such cases are listed in Table~\ref{table_1}, and discussed in Supplementary Material (SM)~\cite{SM}. Interestingly, due to its ${\mathcal T}$-odd nature, $R_{ij}^{l}$ vanishes in magnets that preserve the $t {\mathcal T}$-symmetry, as in many conventional AFM or recently proposed $p$-wave magnets~\cite{yamada_N2025_a,song_N2025_elec}. Moreover, because the magnetic field is even under ${\mathcal P}$, the $R_{ij}^{l}$ is also forbidden in the $ {\mathcal P} {\mathcal T}$-symmetric AFMs as highlighted in Fig.~\ref{fig_1}(c).

{\it Various measurement geometries of linear MR---}
Due to the (third-rank) tensorial nature, the linear MR can manifest in various measurement setups. In the standard resistivity measurement, with $V\parallel I$, one may probe (i) a longitudinal configuration with $I\parallel B$ and (ii) a transverse configuration with $I\perp B$ [see Fig.~\ref{fig_1}(d-e)]. These measurement geometries are relevant for CrSb, Mn$_5$Si$_3$, and KV$_2$Se$_2$O, where linear MR are expected in the channel parallel to the current. 

Complementarily, in Hall-bar–type setups with $V\perp I$, the linear MR signal we discuss is encoded in the symmetric perpendicular resistivity $\tilde R_{ij} = [R_{ij}  + R_{ji}]/2$ and is therefore distinct from the ordinary Hall effect or AHE, which arises from the anti-symmetric part.  Within this perpendicular geometry, one can further distinguish (i) an in-plane field (“planar”) configuration~\cite{goldberg_PRB1954_new}, and (ii) an out-of-plane field configuration [see Fig.~\ref{fig_1}(f-g)]. While several AMs are expected to show the planar response (see Table~\ref{table_1}), AM candidates RuO$_2$, Ca$_3$Ru$_2$O$_7$, and MnTe (spins $\parallel$ $2 \bar 1 \bar 1 0$) only permit out-of-plane symmetric linear MR responses. 

\begin{table*}
\centering
\renewcommand{\arraystretch}{1.8}
\caption{Spin orientations, magnetic space groups (MSG), and allowed conductivities responsible for linear magnetoresistivity (MR) in experimentally realized metallic altermagnets (AMs). The conductivity tensor are defined as $J_i =\sigma_{ij}^{l} E_jB_l$. We have defined $ \bar {\mathcal M}_z \equiv  \{ {\mathcal M}_z |(0, 0,\frac{1}{2}) \}$, $ \bar {\mathcal M}_{y} \equiv \{ {\mathcal M}_{y} |(0, 0,\frac{1}{2}) \} $ for CrSb, $\alpha$-MnTe, and Mn$_5$Si$_3$. The symmetry relating the opposite spin sublattices are highlighted by the column ``AM symmetry" in the presence of spin-orbit coupling, while the symmetries that determine the magnetoconductivity are tabulated in the column ``crucial symmetry". }
\begin{tabular}{|c|c|c|c|c|c|c|}
\hline
Material & Spin axis & MSG & AHE & AM symmetry & Crucial symmetry & ${\mathcal T}$-odd linear MR  \\
\hline

RuO$_2$ [$d$-wave] & 001 & $P4_2^\prime/mnm'$ & $\times$ &
$\{{\mathcal C}_{4z} | (\frac{1}{2}, \frac{1}{2},\frac{1}{2})\} {\mathcal T}$  & ${\mathcal M}_z$, $\{ {\mathcal M}_{x,y} |(\frac{1}{2}, \frac{1}{2},\frac{1}{2}) \} $ &
$\sigma_{xy}^z, \sigma_{yz}^x, \sigma_{xz}^y$  \\
\hline

CrSb [$g$-wave] & 0001 & $P6_3^\prime/m'm'c$ & $\times$ &
$\{{\mathcal C}_{6z} | (0, 0,\frac{1}{2})\} {\mathcal T}$ & $ \bar {\mathcal M}_z  {\mathcal T}$, ${\mathcal M}_x {\mathcal T}$, $\bar  {\mathcal M}_{y}$, ${\mathcal C}_{3z}$ &
$\sigma_{xx}^y,\sigma_{yy}^y,\sigma_{xy}^x$ \\
\hline

$\alpha$-MnTe [$g$-wave]
& $2\bar{1}\bar{1}0$ & $Cmcm$ & $\times$ &
$\{{\mathcal C}_{2y} | (0, 0,\frac{1}{2})\}$ & $\bar {\mathcal M}_z$ , $\bar {\mathcal M}_y$, $ {\mathcal M}_x$ &
{$\sigma_{xy}^z, \sigma_{yz}^x, \sigma_{xz}^y$} \\
\cline{1-7}
Ca$_3$Ru$_2$O$_7$ [$d$-wave] & $010$ & $Pn2_1 a$ & $\times $ & $\{ {\mathcal M}_{x} | (\frac{1}{2},\frac{1}{2}, \frac{1}{2})\} $ &
$ \{ {\mathcal M}_x | (\frac{1}{2},\frac{1}{2}, \frac{1}{2})\} $, $\{ {\mathcal M}_z| (\frac{1}{2}, 0,\frac{1}{2})\}$ & $\sigma_{xy}^z, \sigma_{yz}^x, \sigma_{xz}^y$ \\
\cline{1-7}
Mn$_5$Si$_3$ thin film [$d$-wave] & $1\bar 1 00$ & $Cm^\prime cm^\prime$  & $\times$ & $\{{\mathcal C}_{2z} | (0,0,\frac{1}{2})\} {\mathcal T}$ & $\bar {\mathcal M}_z {\mathcal T}$ , ${\mathcal M}_y$, ${\mathcal M}_x {\mathcal T}$ & $\sigma_{yy}^y$, $\sigma_{xx}^y$, $\sigma_{xy}^x$ \\
\cline{1-7}
KV$_2$Se$_2$O [$d$-wave] & 001 & $P4^\prime /mm^\prime m$ & $\times$ &  ${\mathcal C}_{4z} {\mathcal T}$ &  $ {\mathcal M}_x {\mathcal T} $, ${\mathcal M}_y {\mathcal T}$, ${\mathcal M}_z$, ${\mathcal C}_{4z} {\mathcal T}$ & $\sigma_{xx}^z,\sigma_{yy}^z , \sigma_{zy}^y, \sigma_{zx}^x$
\\
\hline
\end{tabular}
\label{table_1}
\end{table*}

{\it Microscopic theory of linear MR---} Now, we discuss the microscopic origin of the theoretically relevant quantity for linear MR, $\sigma_{ij}^l$, defined as
\be \label{cond}
J_i = \sigma_{ij}^l E_jB_l ~,
\ee
with $J_i$ and $E_j$ as the current density and electric field, respectively. An intuitive understanding of linear conductivity, Eq.~\eqref{cond}, emerges from the steady-state force balance equation~\cite{wang_NC2020_anti}
\be
\frac{d {\bm p}}{d t}= -e {\bm E} -e {\bm p} \times {\bm B} + \alpha {\bm M} \times  {\bm E} - \frac{{\bm p}}{\tau}~.
\ee
The first two terms are the usual Lorentz force, and the last one is momentum relaxation with relaxation time $\tau$. The third term is the Magnus force, which has been used to heuristically explain AHE. It appears in magnetic media and originates from the self-rotating motion, i.e., $\textbf{\textit{M}}$ represents the orbital magnetization. The solution leads to a velocity of the form $({\bm M} \times {\bm E} ) \times {\bm B}= ({\bm M} \cdot {\bm B}) {\bm E} -   ({\bm E} \cdot  {\bm B}) {\bm M} $, which causes the linear MR.

A closely related form emerges in the semiclassical wave-packet theory~\cite{xiao_RMP2010_berry,gao_PRL2014_field}, where Berry curvature $({\bm \Omega})$ plays the role of ${\bm M}$ in the momentum space. Within the Boltzmann transport theory, ${\mathcal T}$-odd magnetoconductivity (MC) is given by~\cite{xiao_PRB2020_linear,SM}
\be \label{berry}
\sigma_{ij, \Omega}^{l} =   - \frac{e^3 \tau}{\hbar}   \int [d{\bm k}] \left[ (v_i  \delta_{jl} + v_j \delta_{il}) {\bm v} \cdot {\bm \Omega} - v_i v_j  \Omega_l \right] f'.
\ee
Here, $-e$ is the electronic charge, $f^\prime$ is the first derivative of the Fermi function with respect to energy, $[d{\bm k}]\equiv d^D{\bm k}/(2 \pi)^D$, and ${\bm v}=1/\hbar {\bm \nabla}_{\bm k} \epsilon$ is the velocity. The first term in Eq.~\eqref{berry} is non-zero only for 3D systems because ${\bm v} \cdot{\bm \Omega} = 0 $ in a 2D system. Furthermore, the first term $v_i \delta_{jl}$ in parentheses originates from the non-orthogonal orientation of fields (${\bm E} \cdot {\bm B}$), which may play a significant role in enhancing the effect in the presence of chiral anomaly in Weyl AMs~\cite{son_PRB2013_chiral}.

In addition to the Berry curvature, the orbital magnetic moment (${\bm m}$) also contributes through the Zeeman coupling. It generates a contribution given by~\cite{xiao_PRB2020_linear,SM}
\bea \label{omm}
\sigma^{l}_{ij, {\rm m}} =  - \frac{e^2 \tau}{\hbar} \int [d{\bm k}]  m_l   \left[  2 \left( \partial_{k_j} v_i \right) f' + \hbar v_i v_j   f^{\prime \prime}  \right] ~.
\eea
The total contribution can then be obtained from the sum $\sigma_{ij}^{l} =\sigma_{ij, \Omega}^{l} + \sigma^{l}_{ij, {\rm m}}$. Importantly, the conductivities are linear in the scattering time, resulting in a $\tau$-independent MR, parametrized by
\be
{\rm MR} = \frac{\sigma_{ij}^l  B_l}{\sigma_{ij}(0)} \propto \frac{e}{\hbar} B {\Omega}_{\rm F}~,
\ee
where $\rm F$ subscript represents the quantities at the Fermi level. This behaviour is fundamentally different from the usual MR that are $\tau$-dependent.

\begin{figure*}[t!]
\centering
\includegraphics[width =\linewidth]{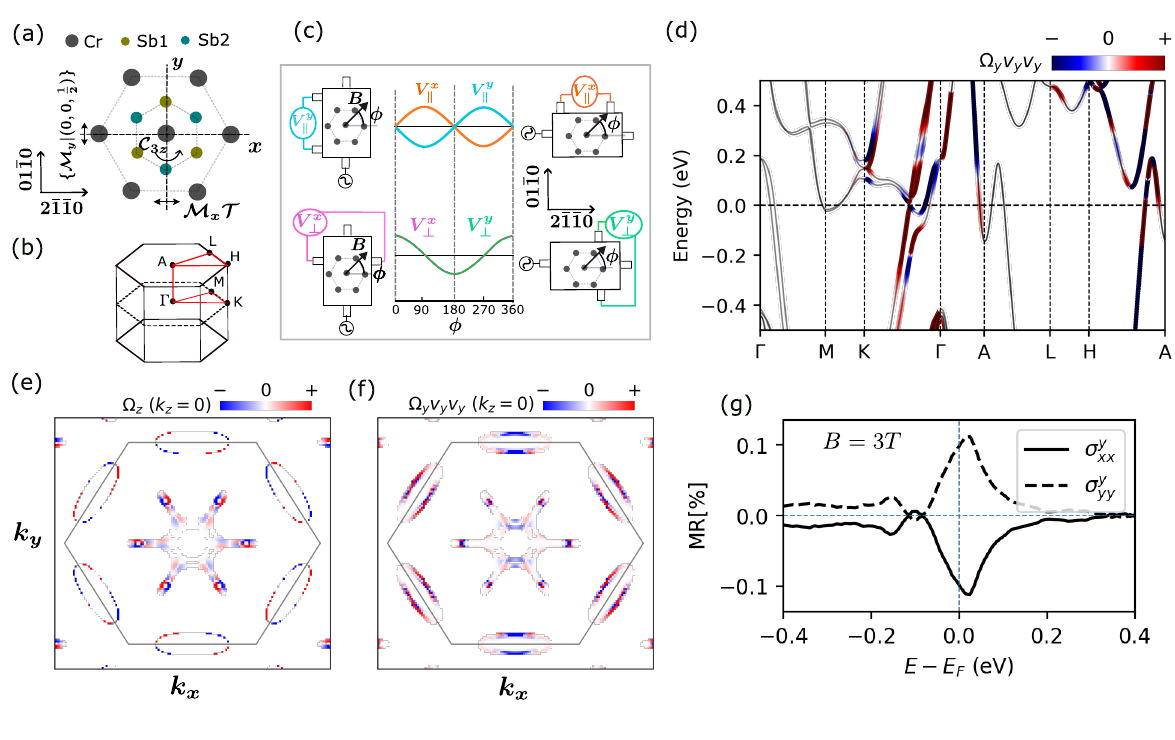}
\caption{{\bf Linear magnetoresistance (MR) in CrSb}: (a) The in-plane crystal structure of CrSb with the ${\mathcal M}_x {\mathcal T}$, $\bar {\mathcal M}_y \equiv \{ {\mathcal M}_y | (0,0,\frac{1}{2})\}$ and ${\mathcal C}_{3z}$ symmetry highlighted. (b) The corresponding hexagonal Brillouin zone (BZ). (c) The expected angular variation of the longitudinal (cyan and orange) and perpendicular (magenta and turquoise) voltage with the magnetic field orientation. (d) The product of Berry curvature and velocity, $ {\mathcal K}_y \equiv \Omega_y v_y v_y$, weighted band structure along the high symmetry line shown in (b). (e) The Berry curvature satisfies $\sum_{n \in \rm occ} \Omega^n_z(k_y) = -\sum_{n \in \rm occ} \Omega^n_z(-k_y)$ respecting the $\bar {\mathcal M}_y$ symmetry leading to vanishing AHE. (f) The product of Berry curvature and velocity in the BZ satisfies $\sum_n {\mathcal K}^n_y (k_y) \delta(\epsilon_n-\epsilon_F)  = \sum_n {\mathcal K}^n_y (-k_y) \delta(\epsilon_n-\epsilon_F)  $, resulting in a finite value of linear MR. (g) The MR percentage at $B=3T$ with the magnetic field along the $y$-axes. }
\label{fig_2}
\end{figure*}

{\it Linear MR in CrSb---} Now, we present the first-principle results of this work focusing on the $g$-wave AM CrSb as a primary case study. CrSb offers an ideal testbed due to its giant non-relativistic spin splitting ($\sim$
1 eV)~\cite{reimers_NC2024_direct,yang_NC2025_three}, and metallic character~\cite{urata_PRM2024_high,peng_PRB2025_scaling}.

In its nonmagnetic phase, CrSb belongs to the hexagonal crystal system with the space group (SG) $P6_3/mmc ~(\#194)$.  Below the transition temperature 710 K, the magnetocrystalline anisotropy forces the spins to align along the 0001 axes, reducing the symmetry to the magnetic SG P$6_3'/m'm'c~(\#194.268)$~\cite{yang_NC2025_three,zhou_N2025_mani}. We have shown the in-plane crystalline symmetry in Fig.~\ref{fig_2}(a). 
We define the spin axes as $z$, in-plane nearest-neighbor (NN) Cr-Cr bonds as $x$-axes. In this convention, the system has six-fold screw symmetry $ \{ {\mathcal C}_{6z} | (0,0,\frac{1}{2}) \} {\mathcal T}$, $z$-axes mirror symmetry $\bar {\mathcal M}_z \equiv \{ {\mathcal M}_z | (0,0,\frac{1}{2})\} {\mathcal T}$, and $y$-axes glide mirror symmetry $\bar {\mathcal M}_y \equiv \{ {\mathcal M}_y | (0,0, \frac{1}{2}) \}$. In addition, the crystal has ${\mathcal M}_x {\mathcal T}$ and ${\mathcal C}_{3z}$ symmetry.

These symmetries constraint the transport coefficient tensors, based on the tranformation properties of current ${\bm J}$ (polar, ${\mathcal T}$-odd), electric field ${\bm E}$ (polar, ${\mathcal T}$-even), and magnetic field ${\bm B}$ (axial, ${\mathcal T}$-odd). For example, under a mirror $\bar {\mathcal M}_y$, polar components perpendicular to the mirror flip sign ($J_y, E_y \to -J_y, -E_y$), while axial components parallel to the mirror flip sign ($B_{x,z} \to -B_{x,z}$). On the other hand, under the anti-unitary symmetry ${\mathcal M}_x\mathcal{T}$, spatial reflection flips the $x$ components, while time reversal reverses ${\bm J}$, ${\bm B}$, but leaves ${\bm E}$ invariant. This yields $J_x, B_{y/z}$ even and $E_x$ odd. Consequently, the $\sigma_{xy}^{\rm A}$ and $\sigma_{yz}^{\rm A}$ components of the anomalous Hall conductivity suppressed by $\bar {\mathcal M}_y$ symmetry. Combined with this, ${\mathcal C}_{3z}$ symmetry further enforces $\sigma_{xz}^{\rm A} = \sigma_{yz}^{\rm A}=0$. 

Applying these rules in Eq.~\eqref{cond}, we find that the ${\mathcal T}$-odd linear MR is allowed only for a magnetic field in the hexagonal plane. A planar MC defined by $\sigma_{xy}^x$ is allowed when the magnetic field is parallel to the NN Cr-Cr bonds. When the magnetic field is along the $y$-axes, transverse MC $\sigma_{xx}^y$ and longitudinal MC $\sigma_{yy}^y$ are allowed. The rest are suppressed. The non-zero components are further related by
\be
\sigma_{xx}^y = \sigma_{xy}^x= -\sigma_{yy}^y ~,
\ee
owing to the ${\mathcal C}_{3z}$ symmetry. Therefore, the standard and planar resistivity will exhibit a $B\sin \phi$ and $B\cos \phi$ dependence, respectively. This is highlighted in Fig.~\ref{fig_2}(c).

We computed the transport coefficients using a Wannier tight-binding Hamiltonian obtained from density functional theory based on experimental lattice parameters~\cite{reimers_NC2024_direct}. See SM~\cite{SM} for details. In Fig.~\ref{fig_2}(d), we project the product of Berry curvature and velocity, ${\mathcal K}_y \equiv \Omega_y v_y v_y$, which appears in $\sigma_{yy}^y$, onto the bands, revealing substantial concentration from the band anti-crossing along the $\Gamma$-K path near the Fermi level. A crucial distinction emerges when comparing this to the AHE. As shown in Fig.~\ref{fig_2}(e), the Berry curvature $\sum_{n \in \rm occu} \Omega^n_{z}$ is odd with respect to $k_y$ (enforced by $\bar {\mathcal M}_y$ symmetry), resulting in exact cancellation upon integration leading to a vanishing $\sigma_{xy}^{\rm A}$. In contrast, $\sum_{n} {\mathcal K}^n_y \delta(\epsilon_n-\epsilon_F)$ is $k_y$ symmetric in momentum space [Fig.~\ref{fig_2}(f)]. Consequently, it survives Brillouin zone integration to yield a finite linear MR shown in Fig.~\ref{fig_2}(g).

%
%

{\it Other AM materials---} Beyond the well-established AM candidates like CrSb, the linear MR may offer a resolution for materials where the AM phase is currently disputed. For instance, there is an ongoing debate regarding the AM phase in RuO$_2$~\cite{hiraishi_PRL2024_nonmag,keler_npjs2024_absence}. Since linear MR is allowed in RuO$_2$ in the as-grown samples (spin parallel to 001), our approach can directly detect ${\mathcal T}$-breaking without reorienting the N\'eel order~\cite{libor_SA2020_crystal,feng_NE2022_an} and remaining within the linear transport regime~\cite{fang_PRL2024_quantum,chu_PRL2025_third}.
Similarly, our approach can help to establish the proposed hidden-order AM state in Ca$_3$Ru$_2$O$_7$~\cite{zhao_PRB2025_nonlinear}, avoiding the need for higher order transport~\cite{mali_NC2026_probing}. Furthermore, recent neutron diffraction measurements in KV$_2$Se$_2$O~\cite{sun_PRB2025_anti} suggest a trivial g-type AFM ground state, in contrast to earlier reports of a metallic AM phase~\cite{jiang_NP2025_a,bai_PRB2024_absence}. Because AHE vanishes when spins align along the easy axes in this system~\cite{yang_PRB2025_magnetic}, the linear MR response predicted here offers a decisive and symmetry-sensitive means of resolving this discrepancy.

{\it Discussions---} Before concluding, several remarks regarding experimental observation and theoretical scope are in order. i) A Néel-vector–dependent resistance based on the interplay of the spin Hall effect and spin splitting effect has been recently proposed~\cite{zhang_PRL2025_theory}. However, this effect does not involve an external magnetic field. In contrast, the linear MR proposed in our work requires a finite magnetic field. ii) The linear MR can be strongly enhanced in Weyl AMs~\cite{antonenko_PRL2025_mirror,li_CP2025_topo,zhao_PRB2025_nonlinear} due to the chiral anomaly~\cite{zyuzin_PRB2017_magneto}. For ${\bm E} \parallel {\bm B}$, the anomaly pumps charge between Weyl nodes of opposite chirality and generates an additional current along the field direction. This contribution is governed by the inter-node scattering time $\tau_\chi$. When inter-node relaxation is slow, i.e., $\tau_\chi \gg \tau$, the anomaly-induced contribution can dominate the MR response. iii) While we focus on orbital coupling of the magnetic field, inclusion of spin-Zeeman coupling~\cite{liu_PRB2025_enhancement,korrapati_PRB2025_aprroxi,ouyang_arxiv2025_quantum} is a natural future direction of our work. 
iv) Experimental care is required regarding magnetic domains, as the predicted linear MR is observable only in the single-domain regime. Domains with opposite Néel orientations cancel the response.

{\it Conclusions---} Our strategy of detecting ${\mathcal T}$-symmetry breaking in AMs via linear MR naturally extends beyond charge transport. The same symmetry principles apply to thermoelectric and thermal responses, where field-linear magneto-thermoelectric effects can probe Néel order~\cite{shtrikman_SSC1965_remarks,yang_NRP2023_the,behnia_RPP2016_nernst}. In insulating AMs, heat transport mediated by phonons or magnons may provide an additional avenue for accessing symmetry-sensitive signatures. Likewise, optical probes can exploit magnetic-field–dependent optical conductivity to reveal hidden AM order in insulating systems~\cite{sunko_arxiv2025_linear} when the MOKE~\cite{pan_PRL2026_experimental} is forbidden. More broadly, this work establishes field-linear transport responses as a versatile symmetry-based diagnostic of unconventional AFMs, complementary to anomalous Hall ~\cite{smejkal_NRP2022_anomalous} and nonlinear magneto-transport~\cite{jiang2025nonlinear,mali_NC2026_probing,fang_PRL2024_quantum}-based probes.

{\it Acknowledgment---} We thank Yufei Zhao for insightful discussions. B.Y. acknowledge the National Science Foundation through the Penn State Materials Research Science and Engineering Center (MRSEC) DMR 2011839 and the Israel Science Foundation (ISF  No. 2974/23).

\appendix

\bibliography{main,refs-altermagnet}

\end{document}